\documentclass[conference]{IEEEtran}
\IEEEoverridecommandlockouts
\usepackage{cite}
\usepackage{amsmath,amssymb,amsfonts}
\usepackage{algorithm}
\usepackage{algorithmic}
\usepackage{graphicx}
\usepackage{textcomp}
\usepackage{xcolor}
\usepackage{caption}
\usepackage{subcaption}
\usepackage{comment}
\usepackage{multirow}
\usepackage{diagbox}
\usepackage{balance}

\usepackage{amsthm}
\newtheorem{observation}{Observation}

\usepackage{enumitem}

\usepackage{color, colortbl}
\definecolor{Gray}{gray}{0.9}
\definecolor{LightCyan}{rgb}{0.88,1,1}

\newcolumntype{a}{>{\columncolor{Gray}}c}
\newcolumntype{b}{>{\columncolor{LightCyan}}c}
\def\BibTeX{{\rm B\kern-.05em{\sc i\kern-.025em b}\kern-.08em
    T\kern-.1667em\lower.7ex\hbox{E}\kern-.125emX}}

\usepackage{pifont}

\begin{document}

\title{Hybrid Workload Scheduling on HPC Systems}

\author{\IEEEauthorblockN{Yuping Fan, Zhiling Lan}
\IEEEauthorblockA{\textit{Illinois Institute of Technology} \\
\textit{Chicago, IL}\\
yfan22@hawk.iit.edu, lan@iit.edu}
\and
\IEEEauthorblockN{Paul Rich, William Allcock}
\IEEEauthorblockA{\textit{Argonne National Laboratory} \\
\textit{Lemont, IL}\\
{\{richp,allcock\}}@anl.gov}
\and
\IEEEauthorblockN{Michael E. Papka}
\IEEEauthorblockA{\textit{Argonne National Laboratory} \\
\textit{Northern Illinois University}\\
papka@anl.gov}
}

\maketitle

\begin{abstract}
Traditionally, on-demand, rigid, and malleable applications have been scheduled and executed on separate systems. The ever-growing workload demands and rapidly developing  HPC infrastructure trigger the interest of converging these applications on a single HPC system. Although allocating the hybrid workloads within one system could potentially improve system efficiency, it is difficult to balance the tradeoff between the responsiveness of on-demand requests, the incentive for malleable jobs, and the performance of rigid applications. In this study, we present several scheduling mechanisms to address the issues involved in co-scheduling on-demand, rigid, and malleable jobs on a single HPC system. We extensively evaluate and compare their performance under various configurations and workloads. Our experimental results show that our proposed mechanisms are capable of serving on-demand workloads with minimal delay, offering incentives for declaring malleability, and improving system performance. 
\end{abstract}


\section{Introduction}
The tremendous compute power with high bandwidth memory and enormous storage capabilities makes high performance computing (HPC) facilities ideal infrastructures for various types of applications. The main tenant of HPC systems is batch applications, which are tightly coupled parallel jobs and are rigid in size. On-demand applications are time-critical applications requiring quick response and thus are used to running on their dedicated clusters. As the sizes of on-demand applications are rapidly expanding in recent years, the dedicated clusters cannot keep up with the rapid expansion in on-demand applications. As a result, HPC system becomes a more practical solution for on-demand applications. Malleable applications are loosely coupled applications consisting of a series of tasks and therefore they can adapt their sizes to changes in hardware availability. Malleable applications are typically running in datacenters. In recent years, an increasing number of HPC systems are equipped with accelerators. The superior computing power combined with the emerging accelerators makes HPC systems an attractive alternative for malleable applications.

The production HPC job schedulers, such as Slurm, Moab/TORQUE, PBS, and Cobalt \cite{SLURM,Moab,PBS,Cobalt}, adopt the traditional batch job scheduling model, where users request a fixed amount of resources for a specific amount of time, while the scheduler decides when and where to run each job based on job priority and system availability. A number of studies attempt to address the hybrid workload scheduling on a single HPC system. Research on co-scheduling rigid and on-demand applications often aims at the high responsiveness of on-demand jobs. The common strategies include predicting on-demand jobs' requests, reserving resources for on-demand jobs, and preempting rigid jobs to make room for on-demand jobs \cite{Liu,Maurya}. Other studies focus on co-scheduling malleable jobs with rigid jobs on HPC systems \cite{Balsam,Carroll,Sun2011,Souza,Chadha,Hungershofer,Prabhakaran}. Unfortunately, none of the existing studies address the problem of co-scheduling all three types of applications, i.e., on-demand jobs, rigid jobs, and malleable jobs. Hence, the scheduling implications of co-running these applications are unknown.

The potential benefits of executing the hybrid workloads on a single HPC system are supporting ever-increasing on-demand job sizes, reducing resource fragmentation, and improving system utilization by malleable jobs. However, this is a challenging task at both job scheduling level and resource management level. The job scheduler needs to maintain the delicate balance between several conflicting objectives, i.e., quick response to on-demand jobs, high system utilization, the incentive for shrinking malleable jobs, and low impact on rigid jobs. The resource manager has to execute the more complicated and frequent operations from job scheduler, i.e., start, preemption, shrink, and expansion operations. This requires more efficient draining processes and corporation with individual applications.

In this paper, we concentrate on addressing HPC hybrid workloads problem from the job scheduling aspect. We propose six mechanisms to address the aforementioned challenges that rely on shrinking, expanding, checkpointing, and resuming techniques to dynamically make room for time-sensitive on-demand applications. Each mechanism focuses on different aspects of scheduling. Our proposed mechanisms are designed to be used in conjunction with the existing scheduling policies: while a scheduling policy determines the order of waiting jobs, our mechanisms manipulate the running jobs in order to meet on-demand requests with minimal impact on other applications. Our design is based on the fact that it is often possible for on-demand jobs to determine their requests within a short time (15-30 minutes) before their actual arrivals \cite{Liu}. Upon receiving on-demand job's advance notice, we provide both non-invasive and invasive mechanisms to reserve resources for the on-demand job. Once an on-demand job arrives, we provide several mechanisms to immediately vacate nodes from running malleable and rigid jobs. By combining the mechanisms used at on-demand job's advance notice and its arrival, we propose six mechanisms to handle hybrid workload scheduling problems. 

To comprehensively evaluate our mechanisms under various scenarios, we conduct a series of trace-based simulations using various workloads generated based on real workload traces collected from Theta \cite{Theta} at Argonne Leadership Computing Facility (ALCF). The results show that all of the proposed mechanisms achieve quick responsiveness for on-demand jobs. Additionally, the results reveal the impact of different mechanisms on system performance and the performance on malleable and rigid jobs. More importantly, we provide valuable insights for choosing these mechanisms under different situations.

\section{Related Work and Challenges}\label{Background and Challenges}
\subsection{HPC Application Types}\label{HPC application types}
\textbf{Rigid job} is the most common type of job in HPC environments \cite{IPPS96}. Rigid jobs have fixed resource requirements throughout their life cycle. Most parallel applications, such as extreme-scale scientific simulations and modeling, are rigid in nature, requiring inter-process communication through message passing, and checkpointing for fault tolerance \cite{Qiao1}. They are tightly coupled applications that cannot be decomposed to a series of small-sized tasks and are prone to failure due to their sizes \cite{Fan8}. In order to handle hardware failures, rigid applications checkpoint regularly and restart from the latest checkpoint in the event of an interruption.

\textbf{On-demand job} is a time-critical application needed to be completed in the shortest time possible. An example of the on-demand jobs is data analytical workloads after experiments \cite{Liu}. Traditionally, to ensure high responsiveness, on-demand jobs are running on dedicated clusters. The rapid experimental expansion requires increasingly large computing capabilities, which cannot be fulfilled by small clusters. The use of large-scale HPC systems becomes a viable solution for the ever-increasing on-demand workloads.

\textbf{Malleable job} is another type of parallel job whose sizes can adapt to the number of nodes assigned to them. 
A malleable job specifies the minimum and the maximum number of nodes. They can shrink down to the minimum sizes or expand up to the maximum sizes based on resource availability. Typically, a malleable job consists of loosely coupled small-sized tasks and the running tasks can be dynamically adjusted based on the assigned nodes. In addition, preemption of malleable jobs causes less overhead than rigid jobs, because they can skip over the finished tasks and resume from the interrupted tasks. The typical examples of malleable jobs are high throughput jobs \cite{Raicu}, multi-task workflows \cite{Balsam}, machine learning applications, and hyperparameter searches in deep neural networks. While traditionally separated infrastructures have been used for rigid jobs and malleable jobs \cite{Sadooghi,He}, the next-generation HPC systems provide not only tremendous compute power on a single node (CPU and GPU), but also enormous high bandwidth memory, making them efficient platforms for malleable workloads \cite{Qiao2}.

\subsection{Job Scheduling in HPC}\label{Job scheduling in HPC}
HPC job scheduling is traditionally designed to manage and assign rigid jobs to resources. The resource allocation is commonly at the granularity of a node. When submitting a job, a user is required to provide job size and job runtime estimate. At each scheduling instance, the scheduler orders the jobs in the queue according to site policies and resource availability and executes jobs from the head of the queue. The most widely used HPC job scheduling policy is First Come First Serve (FCFS) with EASY backfilling \cite{Feitelson02}. FCFS sorts the jobs in the queue according to their arrival times, while backfilling is often used in conjunction with reservation to enhance system utilization. Backfilling allows subsequent jobs in the queue to move ahead under the condition that they do not delay the existing reservations. 

In the realm of executing on-demand jobs and rigid jobs on HPC systems, several groups have proposed to statically or dynamically reserve resources for on-demand requests. Dynamical reservation was achieved by predicting the on-demand request patterns \cite{Liu,Maurya}. In terms of accommodating malleable jobs and rigid jobs on HPC systems, several attempts have been made to shrink malleable jobs in order to reduce resource fragmentation problems \cite{Hungershofer,Prabhakaran,Carroll}. 

To the best of our knowledge, our work is the first attempt to schedule rigid, on-demand, and malleable jobs on a single HPC system. Our design processes on-demand requests by manipulating running rigid and malleable jobs, hence being compatible with existing HPC job scheduling policies concentrating on managing and assigning waiting jobs.

Our work also differs from existing cloud resource managers like Mesos and Kubernetes \cite{Mesos,Kubernetes}. Cloud resource managers commonly allow jobs to share nodes. As the result, solutions for addressing bursty on-demand requests often rely on co-scheduling mixed workloads on a single node, which do not adhere to the constraints of HPC environment, i.e., batch jobs run exclusively on allocated nodes. 

\subsection{Technical Challenges}
Managing hybrid workloads on a single large-scale HPC system offers several potential benefits, such as boosting system utilization, mitigating system fragmentation, and reducing job turnaround time. However, the mixed workloads also pose new challenges.
\begin{itemize}[leftmargin=*]
\item	\textit{Maximize instant start rate of on-demand jobs.} One of the primary goals is to maximize the number of on-demand jobs that can start instantly upon their arrival. By moving the dedicated allocation of on-demand requests to a common resource pool, other types of jobs can utilize these resources and improve system utilization. However, the high system utilization also brings the challenge of the decrease in on-demand instant start rate. 
\item	\textit{Minimize resource waste.} To accommodate the hybrid workloads, a proper mechanism must take advantage of job shrink, expansion, and checkpointing strategies. These strategies come with overheads. For example, to make room for time-critical on-demand requests, we could preempt running rigid/malleable jobs and resume them from the latest checkpoints. These preempted jobs will lose the computation after the checkpoints. Hence, an effective solution must take the resource waste into consideration when choosing running jobs for preemption.
\item	\textit{Incentive of being malleable.} For those jobs that are capable of being adjusted to different sizes, users can either declare them as rigid jobs or malleable jobs. The designed strategies need to provide incentives for users to declare them as malleable jobs by guaranteeing better job performance, i.e., lower average job turnaround time. This could discourage users from lying about their job types. 
\item	\textit{Quick decision making.} To fulfill time-critical on-demand requests, the scheduler has to rapidly choose running jobs to make room for on-demand jobs. Malleable jobs can either be preempted or shrunk, which leads to additional complexity and makes the problem non-trivial. A proper design must be scalable and be capable of making high-quality decisions in a short time (e.g., in seconds).
\end{itemize}

\begin{figure}[htbp]
\centering
\centerline{\includegraphics[width=\linewidth]{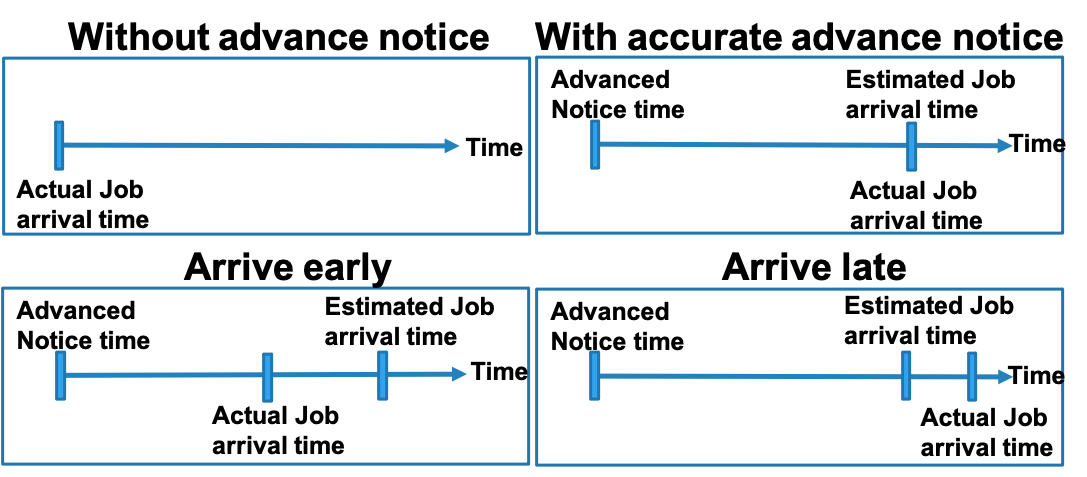}}
\caption{Four types of on-demand jobs.}
\label{advance_notice}
\vspace*{-.5cm}
\end{figure}

\section{Methodology}\label{Methodology}

In this section, we first formally define our hybrid workload scheduling problem in \S\ref{Problem Formulation}. We then present the six scheduling mechanisms to solve this problem in \S\ref{Mechanisms}.

\subsection{Problem Formulation}\label{Problem Formulation}
Suppose an HPC system has $N$ identical nodes. Independent jobs $J_1, J_2, ..., J_n$ arrive and are scheduled in order. Jobs can be classified into three categories:
\begin{itemize}[leftmargin=*]
\item \textbf{Rigid job}: When submitting a rigid job, a user is required to provide two pieces of information: the number of nodes $n$ and job runtime estimate $t_{estimate}$. A rigid job requires a fixed number of nodes, which cannot be adjusted during execution. Job's actual runtime $t_{actual}$ cannot exceed the job runtime estimate ($t_{actual} \leq t_{estimate}$); otherwise, the job will be killed when reaching the runtime estimate \cite{Fan1}. At the beginning of job execution, a job needs some time $t_{setup}$ to set up communication and coordination. During job execution, the job might take regular checkpoints at frequency $t_f$. In case of interruption, the resumed job will first set up communication in $t_{setup}$ time and then resume from the latest checkpoint. As a result, the resumed job will lose the computation between the latest checkpoint and the preempted time.
\item \textbf{On-demand job}: On-demand jobs are time-critical applications, which needed to start within a very short time after submission. On-demand jobs are often possible to determine their resource need within a short time (15-30 minutes) before submission. Advance notice includes the following information: estimated job arrival time, job size, and job runtime estimate. Based on the on-demand job's estimated arrival time and actual arrival time, on-demand jobs can be categorized into four groups as shown in Figure \ref{advance_notice}, i.e., without advance notice, with accurate advance notice, arrive early, and arrive late. 
\item \textbf{Malleable job}: When submitting a malleable job, a user provides the following information: minimum job size $n_{min}$, maximum job size $n_{max}$, job estimate runtime when running at maximum job size $t_{estimate}$. Similar to rigid jobs, we consider setup time at the beginning of the execution. A malleable job is able to run on any integer nodes between minimum job size and maximum job size ($n_{min} \leq n \leq n_{max}$). We assume the linear speedup in addition to the constant setup overhead. Therefore, we can model the job's actual runtime as:
$t_{actual} = t_{single}/n + t_{setup}$.
Here, $t_{single}$ is the application's runtime on a single compute node.
Note that the size of malleable jobs can be adjusted before or during execution according to scheduling policies, which is slightly different from the well-adopted definition of malleable jobs in \cite{IPPS96}. A malleable job typically consists of small-sized tasks and the overhead of changing job size is negligible, and thus it is reasonable to assume no overhead involves in job expansion or shrink. In case of preemption, We adopt Amazon's two minutes warning strategy on spot instance  \cite{AWSWarninng}. The scheduler provides two minutes for malleable jobs to make a checkpoint. The resumed malleable jobs first take $t_{setup}$ to set up and then resume from the previous preempted time. Note that we take the different checkpointing strategies for rigid jobs and malleable jobs. Two minutes is sufficient for a malleable job to store its states to disk and thus it can avoid regular checkpointing.
\end{itemize}

The scheduling problem we study is to allocate resources to on-demand jobs as soon as possible by reserving available nodes and preempting or shrinking running jobs. We aim to increase the responsiveness of on-demand requests with minimal impact on other jobs.

\subsection{Mechanisms}\label{Mechanisms}
We design our hybrid workloads scheduling problem as a series of decisions triggered by \textit{four types of events of on-demand jobs: advance notice, actual arrival, estimated arrival, and completion}. We propose different strategies to handle these events accordingly.
\subsubsection{Advance notice}
The advance notice allows the scheduler to prepare resources for on-demand jobs before their actual arrival. We propose three mechanisms to handle on-demand jobs' advance notice:
\begin{itemize}[leftmargin=*]
\item \textbf{Do nothing (N)}. This is the baseline strategy. The scheduler ignores advance notice and will handle on-demand requests when they actually arrive.
\item	\textbf{Collect-until-actual-arrival (CUA)}. When receiving an on-demand job's advance notice, the scheduler first collects available nodes for this on-demand job. If more nodes are needed, the scheduler collects nodes released by finished jobs until the requested number of nodes is fulfilled or the on-demand job actually arrives. In case of competition from multiple on-demand jobs, the released nodes are assigned to the on-demand job with the earliest advance notice.
\item	\textbf{Collect-until-predicted-arrival (CUP)}. This method first reserves the currently available nodes for the on-demand job. If the on-demand job needs more nodes, this method try to prepare sufficient nodes at its predicted arrival time. First, it collects the nodes that are expected to be released before the on-demand predicted arrival time. Second, it preempts running jobs before the on-demand job's predicted arrival time. 
To minimize the preemption overhead, we try to preempt rigid jobs immediately after checkpointing. If the on-demand job arrives earlier than its predicted arrival time, we stop the preparation and use the strategies in the following subsection to collect more nodes.
\end{itemize}

\begin{figure}[h]
\centerline{\includegraphics[width=\linewidth]{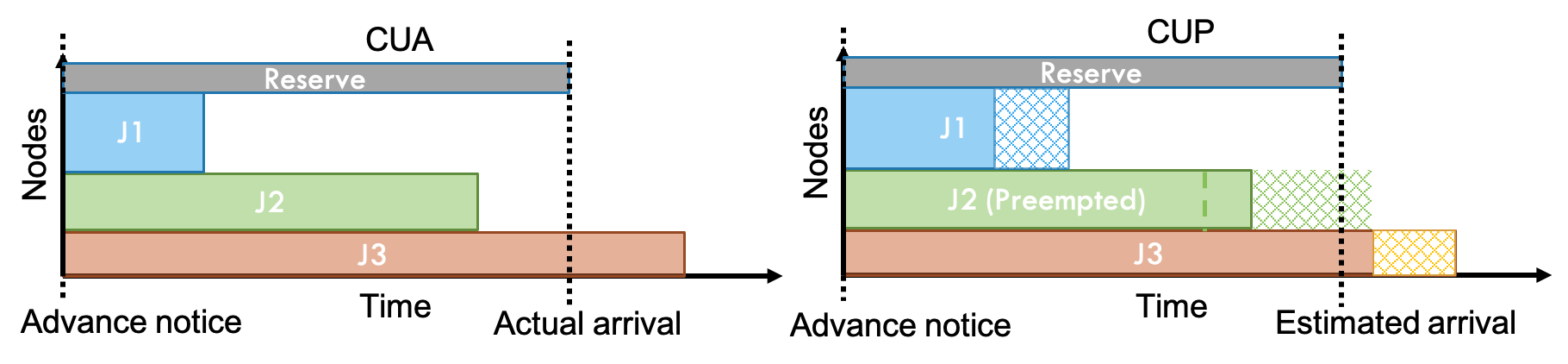}}
\caption{CUA versus CUP. The solid rectangle is the time actually used by a job; the grid rectangle shows the time between job's actual and estimated finish time. CUA reserve the nodes released from J1 and J2. CUP first selects J1; J2 will be preempted immediately after checkpointing (the green dashed line). J2's unfinished computation will be resubmitted and resumed later.}
\label{CUA_CUP}
\vspace*{-.3cm}
\end{figure}

Figure \ref{CUA_CUP} uses an example to illustrate the differences between CUA and CUP. To improve the system utilization, the nodes reserved for on-demand jobs can be used to backfill jobs. But once the on-demand job arrives, all these backfilled jobs have to be preempted immediately.

\subsubsection{On-demand job's actual arrival}
When an on-demand job arrives, the scheduler first checks if there are sufficient available nodes and reserved nodes to run this job. If that is the case, the on-demand job can launch immediately. Otherwise, we propose two strategies to find more nodes:
\begin{itemize}[leftmargin=*]
\item	\textbf{Preempt-at-actual-arrival (PAA)}: This method lists all currently running malleable and rigid jobs in ascending order of their preemption overheads. If the total number of the preemptable nodes is not sufficient, we cannot start the on-demand job instantly and have to put it to the front of the queue waiting for additional available nodes. If the preemptable nodes are sufficient, we preempt jobs from the front of the running list until the on-demand request is satisfied. We update the preempted jobs' estimated runtime, keep their original submit time, and automatically resubmit these jobs to the wait queue. The priority of the preempted jobs is determined by the scheduling policy. For example, FCFS might move the preempted jobs to the front of the queue, because they have very early first submission times.
\item	\textbf{Shrink-preempt-at-actual-arrival (SPAA)}: This method first finds all currently running malleable jobs and computes the maximum number of nodes they can supply by shrinking to their minimum sizes. If the supply can meet the on-demand job's request, the running malleable jobs will shrink their sizes evenly and linearly adjusts their estimated runtimes. If the supply cannot meet, we will use PAA method to handle the on-demand request.
\end{itemize}

\subsubsection{Completion of on-demand job}
For job fairness, once an on-demand job is completed, the on-demand job will try to return its nodes to the lenders. If a job was preempted by this on-demand job and is still waiting in the queue, the leased nodes will return to this job and this job will resume immediately if possible. If a job was shrunk to make room for the on-demand job and the job is still running, we will expand this job to its original size.

\subsubsection{On-demand job's estimated arrival}
An on-demand job may arrive late or even do not show up. To preempt deadlock, if an on-demand job has not arrived after a certain period of time of its estimated arrival time, the scheduler will release the reserved nodes.

By combining three advance notice strategies with two job arrival strategies, we obtain six mechanisms to schedule hybrid workloads on a single HPC system: \textit{N\&PAA, N\&SPAA, CUA\&PAA, CUA\&SPAA, CUP\&PAA, CUP\&SPAA}.

\section{Experimental Setup}\label{Experimental Setup}

In this section, we first describe the real workload trace collected from Theta and how to generate traces from the real trace to represent various scenarios (\S\ref{Workloads}). We then introduce the baseline configuration for our experiments (\S\ref{Configuration}) and our simulation environment (\S\ref{Simulation Platform}). Finally, we list the system- and user-centric metrics for evaluation (\S\ref{Evaluation Metrics}).

\begin{table}[ht]
\begin{minipage}[b]{0.5\linewidth}
\centering
\resizebox{\linewidth}{!}{
\begin{tabular}{|l|l|l|}
\hline
\rowcolor{LightCyan}
                              &Theta                                                                         \\ \hline
\rowcolor{Gray}
Location                      &ALCF                                                                                            \\ \hline
\rowcolor{LightCyan}
Scheduler             & Cobalt                                                                                             \\ \hline
\rowcolor{Gray}
Compute Nodes                          & \begin{tabular}[c]{@{}l@{}}4,392 KNL\end{tabular} \\ \hline
\rowcolor{LightCyan}
Trace Period                  & Jan. - Dec. 2019                                                                       \\ \hline
\rowcolor{Gray}
Number of Jobs & 37,298                                                                                 \\ \hline
\rowcolor{LightCyan}
Number of Projects & 211                                                                            \\ \hline
\rowcolor{Gray}
Maximum Job Length &1 day                                                                       \\ \hline
\rowcolor{LightCyan}
Minimum Job Size &128 nodes                                                                     \\ \hline
\end{tabular}
}
\caption{Theta workload.}
\label{Theta_workload_table}
\end{minipage}\hfill
\begin{minipage}[b]{0.5\linewidth}
\centering
\includegraphics[width=\textwidth]{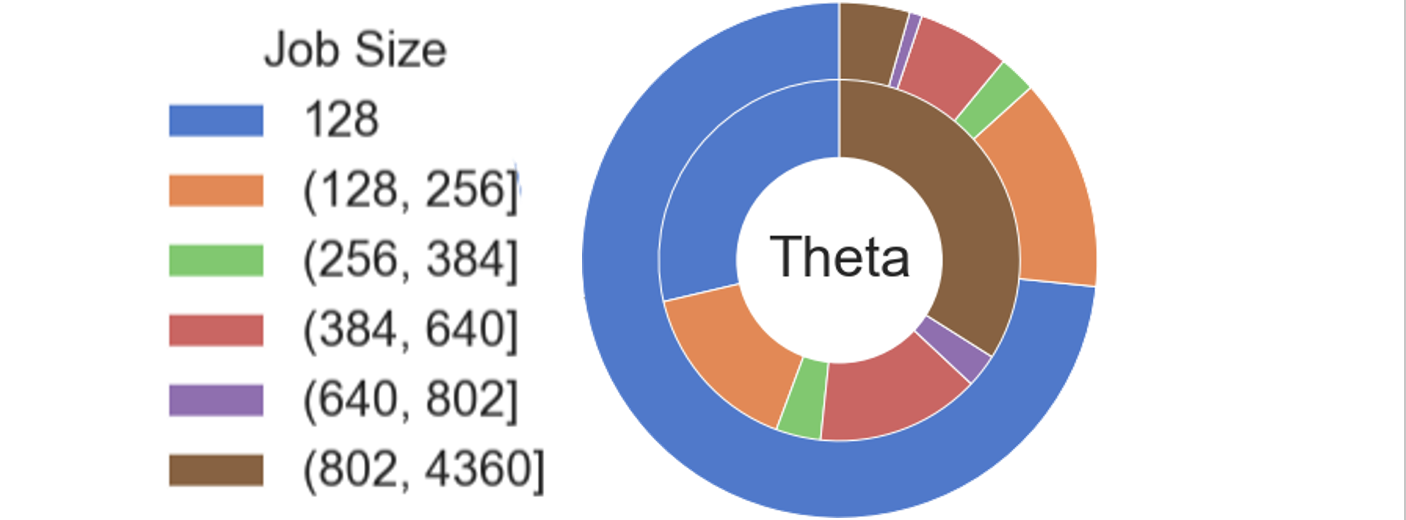}
\captionof{figure}{The number of jobs (outer) and the total core hours (inner) in different size ranges.}
\label{Theta_workload_fig}
\end{minipage}
\end{table}

\begin{table}[ht]
\begin{minipage}[b]{0.5\linewidth}
\centering
\includegraphics[width=\textwidth]{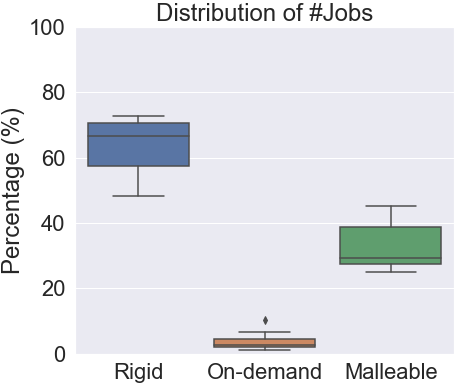}
\captionof{figure}{Job type distributions on the traces in our experiments.}
\label{job_distribution}
\end{minipage}\hfill
\begin{minipage}[b]{0.48\linewidth}
\centering
\includegraphics[width=\textwidth]{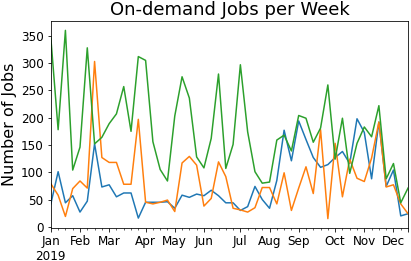}
\captionof{figure}{The number of on-demand jobs per week. Three sample traces, denoted by blue, orange, and green lines, show their bursty behavior.}
\label{on_demand_requests}
\end{minipage}
\vspace*{-0.2in}
\end{table}

\subsection{Workloads}\label{Workloads}
In this study, we use the one-year workload trace from Theta. Theta is a Cray XC40 machine located at ALCF consisting of 4392 compute nodes with 11.69PF peak performance. 
Table I  summarizes the basic information and Figure 3 presents job characterization of Theta workload. 
The workload trace includes basic information for scheduling, such as job submission time, runtime estimate, size, username, and project name. Since the trace does not include job type information, we generate a series of workloads based on the real trace to cover various job distributions. 
Studies have been shown that real on-demand jobs are relatively small in size and have burst submission patterns \cite{Liu}. Users tend to submit a bunch of on-demand jobs in a short period of time. In order to mimic the bursty on-demand job submission pattern, we group jobs by their project names and assume that all jobs belonging to one project have the same job types (i.e., rigid, on-demand, or malleable jobs). For those large on-demand jobs (i.e., job size larger than half of the system size), we randomly reassign them to be rigid jobs or malleable jobs. 

Figure \ref{job_distribution} shows the statistics of traces used in our experiments. We observe that the rigid, on-demand, malleable job distributions differ significantly on different traces because different projects have significant differences in sizes and submission patterns. 
Figure \ref{on_demand_requests} shows the on-demand requests of three sample traces. The traces show the bursty on-demand job submission pattern. This enables us to extensively evaluate our mechanisms under various scenarios. 

\subsection{Configuration}\label{Configuration}
In this section, we present the default configurations for different types of jobs. By default, waiting jobs are scheduled by FCFS with EASY backfilling. We assign that 10\% of projects submit on-demand jobs, 60\% of projects submit rigid jobs and the rest of projects submit malleable jobs. We made this assumption because rigid jobs are the main tenant and malleable jobs are emerging in HPC systems. HPC systems can support limited on-demand requests to ensure their responsiveness. 

In terms of rigid jobs, we set their setup overhead to be 5\%-10\% of their runtimes. We assume rigid jobs make regular checkpoints at the optimal frequency defined by Daly \cite{Daly}. Based on our experience
and the current literature \cite{Topper,Moody}, we set each checkpointing overhead to 600 seconds if the job used less than 1K nodes; otherwise, we set it to 1200 seconds. 

In terms of on-demand jobs, we equally distribute them into the following categories: without advance notice, with accurate advance notice, arrive early, and arrive late. If an on-demand job arrives early, its arrival time is a random number between its advance notice and estimated arrival time. If an on-demand job arrives late, its arrival time is a random number within 30 minutes after its estimated arrival time. We set the threshold to release the reserved nodes to 10 minutes after the on-demand job's estimated arrival time.

In terms of malleable jobs, we set their maximum job size to be their original requested job size and their minimum job size to be 20\% of their maximum size. The setup overhead is a random number between 0\%-5\% of their runtimes.


\subsection{Trace-based Simulation}\label{Simulation Platform}
We compare different scheduling mechanisms through trace-based simulation. Specifically, a trace-based, event-driven scheduling simulator called CQSim is used in our experiments \cite{xu01,Fan4,Li1,Fan5,CQSimGithub,Fan6,Fan7}. CQSim contains a queue manager and a scheduler that can plug in different scheduling policies. It emulates the actual scheduling environment. A real system takes jobs from user submission, while CQSim takes jobs by reading the job arrival information in the trace. Rather than executing jobs on system, CQSim simulates the execution by advancing the simulation clock according to the job runtime information in the trace. 

\subsection{Evaluation Metrics}\label{Evaluation Metrics}
We evaluate the performance of different mechanisms using several user-level and system-level metrics.
\begin{enumerate}[leftmargin=*]
\item	\textbf{Job turnaround time} is a user-level metric. It measures the interval between job submission and completion time.
\item \textbf{On-demand jobs' instant start rate} is a user-level metric, which is calculated as the ratio between the number of on-demand jobs started instantly and the total number of on-demand jobs. 
\item	\textbf{Preemption ratio} is a user-level metric to measure the percentage of rigid or malleable jobs being preempted.
\item	\textbf{System utilization} is a system-level metric that measures the ratio of node-hours used for useful job execution to the total elapsed node-hours. Note that system utilization excludes wasted computation due to preemption.
\end{enumerate}

\begin{figure*}[!h]
 \centering
\begin{subfigure}{\textwidth}
  \includegraphics[width=\linewidth]{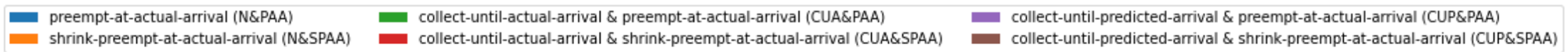}  
\end{subfigure}
\begin{subfigure}{0.327\textwidth}
  \centering
  \includegraphics[width=\linewidth]{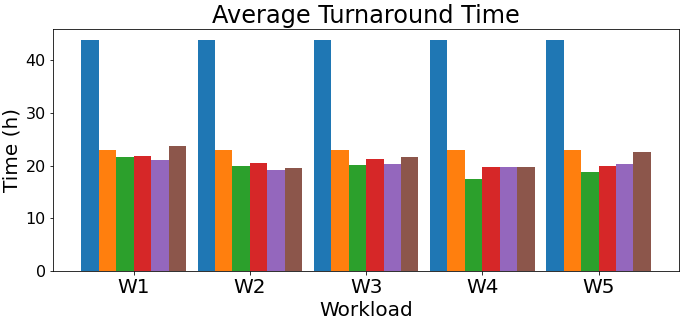} 
\end{subfigure}
\begin{subfigure}{0.327\textwidth}
  \centering
  \includegraphics[width=\linewidth]{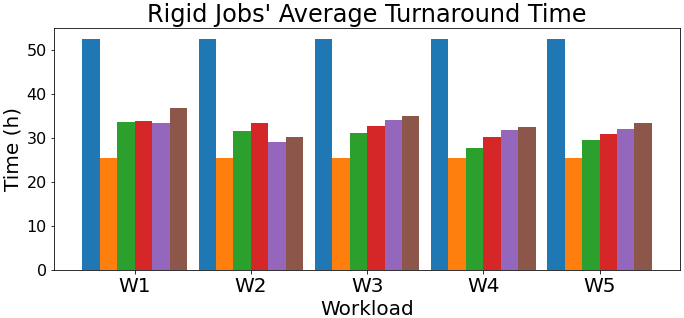}  
\end{subfigure}
\begin{subfigure}{0.327\textwidth}
  \centering
  \includegraphics[width=\linewidth]{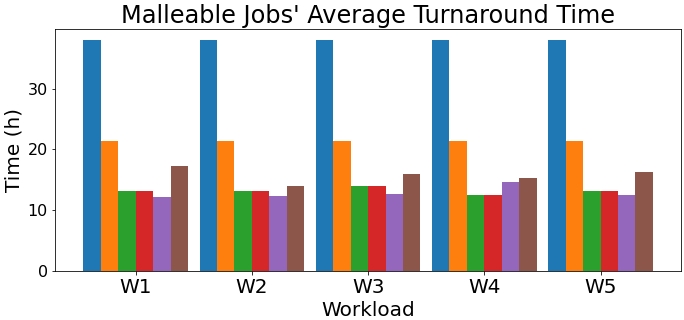}  
\end{subfigure}
\begin{subfigure}{0.245\textwidth}
  \centering
  \includegraphics[width=\linewidth]{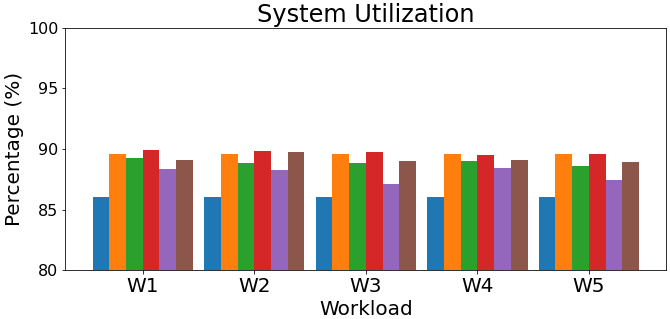}  
\end{subfigure}
\begin{subfigure}{0.245\textwidth}
  \centering
  \includegraphics[width=\linewidth]{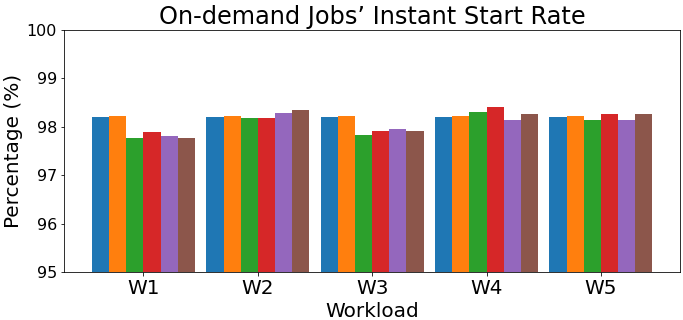}  
\end{subfigure}
\begin{subfigure}{0.245\textwidth}
  \centering
  \includegraphics[width=\linewidth]{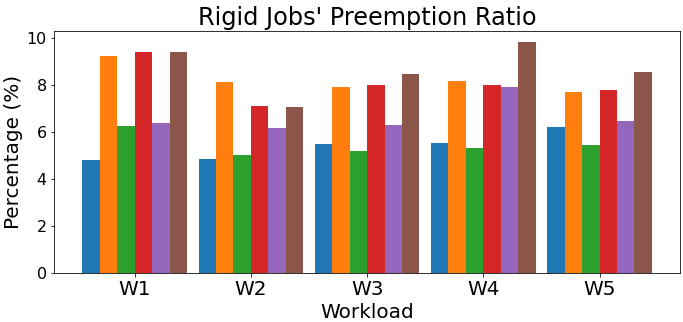}  
\end{subfigure}
\begin{subfigure}{0.245\textwidth}
  \centering
  \includegraphics[width=\linewidth]{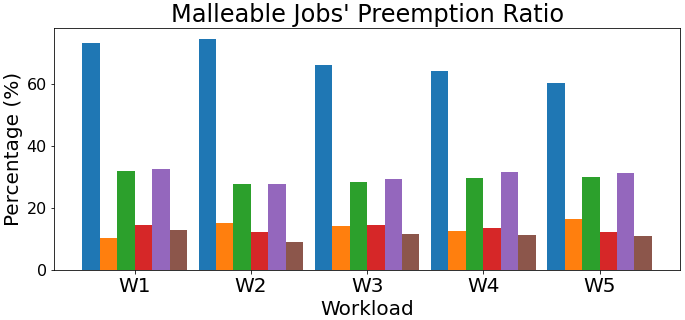}  
\end{subfigure}
\caption{Scheduling performance under different advance notice accuracies (shown in Table \ref{advance_notice_workload_distribution}). To show the performance under various situations, we repeat the same experiment on ten randomly generated traces and the results in this section are averaged.}
\label{advance_notice_results}
\vspace*{-.4cm}
\end{figure*}

\section{Evaluation}\label{Evaluation}
To comprehensively evaluate the six proposed mechanisms, we conduct a series of experiments to compare them under various situations/configurations, including different advance notice settings (\S\ref{Impact of the Accuracy of Advance Notice}), and different checkpoint settings (\S\ref{Impact of Checkpointing Frequency}).

\begin{table}[h]
\centering
\caption{Baseline performance. Baseline algorithm is FCFS/EASY backfilling without special treatments on on-demand, rigid, and malleable jobs.}
\label{baseline_performance}
\begin{tabular}{|l|l|l|}
\hline
\rowcolor{LightCyan}
   Avg. Turnaround & System Util. & On-demand Jobs' Instant Start Rate\\ \hline
\rowcolor{Gray}
15.6 hours      & 83.93\%            & 22.69\%         \\ \hline
\end{tabular}
\vspace*{-.4cm}
\end{table}

\begin{table}[h]
\centering
\caption{Distribution of on-demand jobs in different workloads. Take W1 as an example: 70\% of on-demand jobs arrive without advance notice; 10\% of on-demand jobs arrive with accurate advance notice; 10\% of on-demand jobs arrive early; the rest 10\% of on-demand jobs arrive late.}
\label{advance_notice_workload_distribution}
\begin{tabular}{|l|l|l|l|l|}
\hline
\rowcolor{LightCyan}
   & No Notice & Accurate Notice & Arrive Early & Arrive Late \\ \hline
\rowcolor{Gray}
W1 & 70\%      & 10\%            & 10\%         & 10\%        \\ \hline
\rowcolor{LightCyan}
W2 & 10\%      & 70\%            & 10\%         & 10\%        \\ \hline
\rowcolor{Gray}
W3 & 10\%      & 10\%            & 70\%         & 10\%        \\ \hline
\rowcolor{LightCyan}
W4 & 10\%      & 10\%            & 10\%         & 70\%        \\ \hline
\rowcolor{Gray}
W5 & 25\%      & 25\%            & 25\%         & 25\%        \\ \hline
\end{tabular}
\vspace*{-.4cm}
\end{table}

\subsection{Overall Performance}\label{Overall Performance}

Figure \ref{advance_notice_results} compares the performance of the five types of workloads, as shown in Table \ref{advance_notice_workload_distribution}, under different on-demand request accuracies. 
In this subsection, we make several interesting observations on overall performance from Figure \ref{advance_notice_results}. 
In the next subsection, we will analyze the impact of advance notice accuracies using these figures.


\begin{observation}
Comparing with FCFS/EASY, the proposed methods boost system utilization and on-demand jobs' instant start rate, while slightly increase average job turnaround time.
\end{observation}

Comparing results in Table \ref{baseline_performance} and Figure \ref{advance_notice_results}, we observe that the proposed methods improves system utilization up to 90\%. The on-demand jobs' instant start rate increases from 22\% to 98\%. The average job turnaround time increases from 15 hours to 22 hours, due to job preemption and shrink.

\begin{observation}
N\&PAA has the worst overall performance.
\end{observation}

N\&PAA obtains the worst results on average job turnaround time and system utilization. Additionally, its malleable jobs' preemption ratio is noticeably higher than other mechanisms. It suggests that this method has a higher preemption overhead and wastes more computation cycles than other methods. The high average job turnaround time is caused by the starvation of some high-priority jobs. Here, starvation means that jobs were preempted, but could not resume for a long period of time after preemption. Although on-demand jobs return their leased nodes to the lenders, the lenders might not resume immediately, because that on-demand jobs might need a portion of the preempted nodes and the rest are moved to the common resource pool. When the on-demand job is finished, the preempted job can only reclaim the nodes from the on-demand job and it has to wait until more nodes are available.  

\begin{observation}
To achieve higher system utilization and lower malleable jobs' preemption ratio, SPAA methods are preferred than PAA methods.
\end{observation}

All three SPAA methods largely reduce malleable jobs' preemption ratio, while slightly increase the rigid jobs' preemption ratio. This is because SPAA attempts to find shrink options, which reduces malleable jobs' preemption ratio. Shrink, in general, has lower overhead and leads to fewer wasted computation cycles and therefore higher system utilization. 

\begin{observation}
To obtain lower average job turnaround time and lower rigid jobs' preemption ratio, PAA methods are recommended than SPAA methods, except N\&PAA method.
\end{observation}

In general, SPAA methods tend to prolong average job turnaround time, especially malleable jobs, because it reduces all running malleable jobs' sizes and prolongs their execution time. On the other hand, PAA affects fewer running jobs and the preempted jobs might resume when the on-demand job finishes. However, N\&PAA is an exception. This is because CUA and CUP prepare some nodes for on-demand jobs before their arrival and PAA only needs to preempt small-sized running jobs upon on-demand job arrival. On the other hand, N\&PAA is more likely to preempt large-sized running jobs, which are more difficult to reclaim their preempted nodes.

It is interesting to notice that PAA methods lead to slightly lower rigid jobs' preemption ratio than SPAA methods. Since SPAA methods first try to shrink malleable jobs, the job sizes of running jobs are, on average, smaller than that of PAA methods. When the shrink option is not possible, SPAA methods need to preempt more running jobs, which causes the slight increases in the rigid jobs' preemption ratio.

\begin{figure*}[!h]
 \centering
\begin{subfigure}{\textwidth}
  \includegraphics[width=\linewidth]{fig/legend.png}  
\end{subfigure}
\begin{subfigure}{0.3\textwidth}
  \centering
  \includegraphics[width=\linewidth]{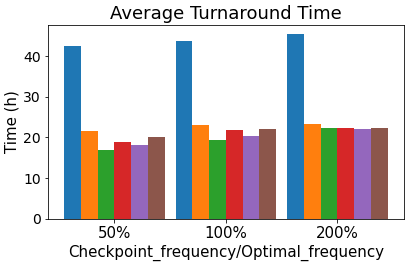} 
\end{subfigure}
\begin{subfigure}{0.3\textwidth}
  \centering
  \includegraphics[width=\linewidth]{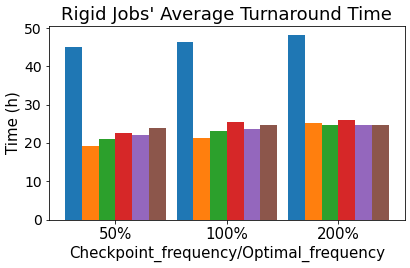}  
\end{subfigure}
\begin{subfigure}{0.3\textwidth}
  \centering
  \includegraphics[width=\linewidth]{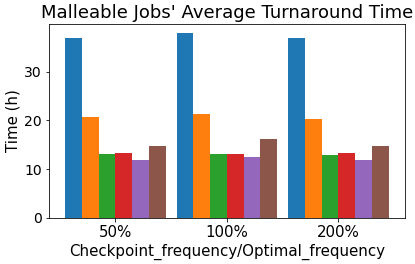}  
\end{subfigure}
\begin{subfigure}{0.23\textwidth}
  \centering
  \includegraphics[width=\linewidth]{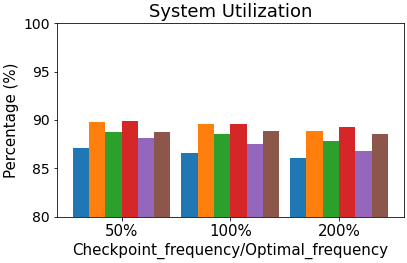}  
\end{subfigure}
\begin{subfigure}{0.23\textwidth}
  \centering
  \includegraphics[width=\linewidth]{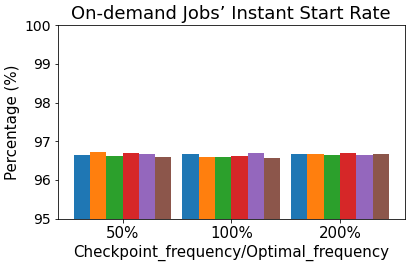}  
\end{subfigure}
\begin{subfigure}{0.23\textwidth}
  \centering
  \includegraphics[width=\linewidth]{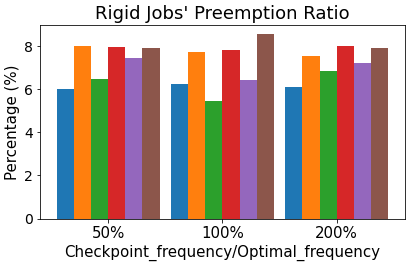}  
\end{subfigure}
\begin{subfigure}{0.23\textwidth}
  \centering
  \includegraphics[width=\linewidth]{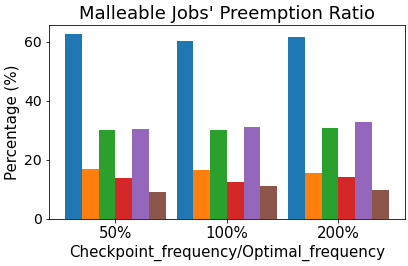}  
\end{subfigure}
\caption{Impact of rigid jobs' checkpointing frequency on scheduling performance. 50\% means rigid jobs makes checkpoints twice as frequent as the optimal checkpointing frequency. }
\vspace*{-0.5cm}
\label{checkpoint_results}
\end{figure*}

\begin{observation}
CUA methods, in most cases, perform better than CUP methods.
\end{observation}

As shown in Figure \ref{advance_notice_results}, CUA methods achieve slightly lower average job turnaround time and slightly higher system utilization. Recall that CUA methods passively collect released nodes, while CUP methods might need to preempt some running jobs before on-demand job arrival. CUA methods trigger fewer preemptions, leading to less resource waste and higher system utilization.

\begin{observation}
CUA\&PAA, CUA\&SPAA, CUP\&PAA, and CUP\&SPAA encourage users to honestly declare their malleable jobs.
\end{observation}

The malleable jobs' turnaround time of these four methods is noticeably lower than the rigid jobs' turnaround time. For all proposed mechanisms, the scheduler can choose malleable jobs' sizes at their start or resumed time. For SPAA methods, although malleable jobs might need to shrink their sizes upon arrival of on-demand jobs, they are guaranteed to expand to their original sizes by reclaiming their released nodes when the on-demand job finishes. The malleability feature significantly increases the chances of malleable jobs being chosen to execute, leading to lower average turnaround time compared with rigid jobs. The better job performance on malleable jobs discourages users from declaring malleable jobs as rigid jobs.

\begin{observation}
N\&SPAA method is a good option when the rigid jobs need to achieve low average turnaround time.
\end{observation}

N\&SPAA achieves the lowest rigid jobs' average turnaround time among the six methods. More importantly, both of its rigid jobs' average turnaround time and malleable jobs' average turnaround time are around 24 hours. When an on-demand job arrives, N\&SPAA first attempts to find shrink options. If there are viable shrink options, the selected malleable jobs will be shrunk and prolonged, while running rigid jobs are not impacted. Upon on-demand job arrival, N\&SPAA requests more nodes than CUA\&SPAA and CUP\&SPAA. Therefore, N\&SPAA has more noticeable adverse effects on malleable jobs than the other SPAA methods. Although N\&SPAA does not provide strong incentives for malleable jobs, it might be a good option for system administrators when rigid jobs have higher priority than malleable jobs.

\begin{observation}
Malleable jobs' preemption ratio is noticeably higher than rigid jobs' preemption ratio.
\end{observation}

This is due to the fact that the preemption overheads of malleable jobs are lower than rigid jobs. In order to minimize wasted computation cycles caused by preemption, the running jobs are preempted in ascending order of their preemption overheads. Malleable jobs only waste their setup times. On the other hand, rigid jobs not only waste their setup times but also lose the computation after the latest checkpoints. It is interesting to notice that despite the higher preemption ratio, malleable jobs achieve lower average turnaround times because they are more likely to run by shrinking their sizes. 

\begin{observation}
All methods achieve extremely high on-demand jobs' instant start rate. 
\end{observation}

As shown in Figure \ref{job_distribution}, on-demand jobs represent 3\%-15\% of total workloads. On average, 98\% of on-demand jobs start instantly. There is no significant difference in on-demand jobs' instant start rate between the different methods. The reason why an on-demand job fails to start immediately is that the nodes used by currently running on-demand jobs plus this on-demand job exceed the whole system capacity. This metric is more related to the on-demand jobs' submission pattern. Bursty on-demand job submission pattern could negatively affect the on-demand jobs' instant start rate.

\begin{observation}
The proposed mechanisms pose trivial scheduling overheads.
\end{observation}

Current HPC systems typically require a scheduler to respond in 10-30 seconds \cite{Fan3,Fan2}. In our experiments, the proposed methods take less than 10 milliseconds to make a decision, hence being feasible for online deployment.

\subsection{Impact of Accuracy of Advance Notice}\label{Impact of the Accuracy of Advance Notice}
\begin{observation}
The performance of CUP methods highly relies on accuracy of advance notice. The more accurate advance notice, the better performance.
\end{observation}

CUP\&PAA and CUP\&SPAA methods achieve their best performance on W2, i.e., the workloads with the highest percentage of on-demand jobs with accurate advance notice. The accurate advance notice reduces the preemption ratio on both rigid and malleable jobs and therefore reduces wasted cycles and improves system utilization. The accurate advance notice also reduces average job turnaround time due to less interruption during execution.

\begin{observation}
The earlier the advance notice, the better the performance of CUA methods. 
\end{observation}

CUA methods obtain the lowest average job turnaround time on W4, i.e., the workloads with the majority of on-demand jobs arrived late. W4 provides a longer period of time between advance notice and job actual arrival. As a result, CUA methods are more likely to collect nodes before the actual arrival of on-demand jobs, and thus decrease the chances of preempting or shrinking running jobs upon arrival of the on-demand jobs. In addition, by preparing more nodes for on-demand jobs before their arrival, it also slightly improves on-demand jobs' instant start rate.

\subsection{Impact of Checkpointing Frequency}\label{Impact of Checkpointing Frequency}

Figure \ref{checkpoint_results} presents the scheduling results under different checkpointing frequencies.

\begin{observation}
To achieve better rigid job performance and system performance, we suggest that rigid jobs take more frequent checkpoints than the optimal checkpointing frequency.
\end{observation}

All methods benefit from the more frequent checkpointing frequency. More frequent checkpoints can reduce rigid jobs' turnaround time and also improve system utilization. Daly's optimal checkpointing frequency is designed for fault tolerance \cite{Daly}. However, the interruption caused by failures is obviously much lower than preemption caused by draining nodes for on-demand jobs. Therefore, increasing checkpointing frequency reduces rigid jobs' lost computation and thus reduces their turnaround time. This also helps improve system utilization by reducing preemption overheads.

\section{Conclusion}\label{Conclusion}
In this paper, we have defined and modeled HPC hybrid workload scheduling problem within one HPC system. We have proposed several mechanisms to reconcile the demands from on-demand, rigid, and malleable applications. By exploring how different mechanisms behave under various configurations and workloads, we have found that our proposed mechanisms enable a single HPC system to accommodate on-demand jobs with minimal delay, encourage users to honestly declare their malleable applications, and minimize overheads caused by preemption. Additionally, we have provided several key insights on different mechanisms. In particular, upon on-demand job's advance notice, CUA performs better than CUP. SPAA is, in general, preferred over PAA upon on-demand job arrival. Finally, N\&SPAA achieves surprisingly good performance on rigid jobs.

\section*{Acknowledgment}
This work is supported in part by US National Science Foundation grants CNS-1717763, CCF-1618776 and U.S. Department of Energy, Office of Science, under contract DE-AC02-06CH11357. 

\balance
\bibliographystyle{unsrt}
\bibliography{ipdps1.bib}

\end{document}